\documentclass[fleqn,10pt]{wlscirep}
\usepackage[utf8]{inputenc}
\usepackage[T1]{fontenc}

\usepackage{physics}
\usepackage{comment}

\usepackage{caption}
\usepackage[subrefformat=parens]{subcaption}

\usepackage{afterpage}

\usepackage{xr}
\title{Characterizing limit order books in call auctions of a stock market}

\author[1]{Shota Nagumo}
\author[1,2]{Takashi Shimada}
\affil[1]{Department of Systems Innovation,
Graduate School of Engineering,
The University of Tokyo, 
Tokyo, 113-8656, Japan}
\affil[2]{Mathematics and Informatics Center, The University of Tokyo,
Tokyo, 113-8656, Japan}

\begin{abstract}
Statistical and dynamical characters of stock markets have been extensively studied, which now is providing the firm basis for econophysics and its application as ``stylized facts''. However, most of those studies are for markets under the continuous auction, i.e. trades are executed sequentially. There has been less research on another major type of auction, call auctions, where orders are accumulated and those are executed at once in the final moment.
This study focuses on the structure of the limit order books of stocks under the call auctions.
Using the data of all stocks listed in the Tokyo Stock Exchange, we find that the shape of the limit order books in call auctions are well fitted by a simple functional form of hyperbolic tangent.
From the fitting, we define the ``median spread'' and the ``width'' of limit orders. The ratio of the ``width'' to the ``median spread'' of most stocks are found to be similar, indicating that the execution ratio (the trading volume relative to the total number of orders) are nearly equal among them. Furthermore, the deviation in this ratio from the majority is found to be a good indicator for finding the stocks of the companies making outstanding profit. Our results demonstrate that those parameters of the structure of the limit order book well characterizes the states of the market under call auctions.

\end{abstract}
\begin{document}

\flushbottom
\maketitle
%
%
\thispagestyle{empty}

\section*{Introduction}

A limit order book is a list of orders to sell or buy at a certain price (limit orders). 
The execution prices and the trading volumes are determined by the structure of the limit order book and the inflow of the market orders (orders which do not specify the price). Therefore, understanding the distribution of the limit order book and the dynamics of that is essential to predicting the dynamics of price, trading volume, volatility, etc. of that stock~\cite{chakraborti2011econophysics1,chakraborti2011econophysics2,bouchaud2009markets,bouchaud2018trades,Abergel2016Limit,gould2013limit}.

The classical indicators to characterize the shape of the limit order books are ``{\it spread}'', ``{\it depth}'', and ``{\it resiliency}''~\cite{kyle1985continuous}.
The {\it spread}  refers to  the distance between sell orders and buy orders. 
The most commonly used one is the Best Bid-Offer spread (BBO spread), which is the distance between  the best ask (the lowest sell orders) and the best bid (the highest buy orders).
 It is known that stocks of the company with lower profit tend to have wider BBO spreads. 
One possible explanation for this is  that  traders demand a high return for taking the risk, and hence place high sell orders and low buy orders~\cite{agrawal2004bid}.
The {\it depth} characterizes the volume of orders near the price of the best ask and the best bid.
Especially,  order volume at the price of the best ask and the best bid is commonly used.
The {\it resiliency}  refers to the speed with which the market price (especially, the BBO spread) returns to its original value after the execution.
Another characteristic of the shape of the limit order books is  `{`\it slope}'', which is the increment of the order volume to the increment of the price on the limit order books.
It is known that 
the slope of the limit order book is negatively correlated with the trading volume~\cite{naes2006order}.

However, these studies focus on the continuous auction, where orders are placed and executed sequentially.
There has been little study on the call auction, where orders are accumulated and those are executed all at once at the end of the auction.
The share of the call auction is at around 30\% in the Tokyo Stock Exchange~\cite{lehmann1994trading, Ohta2008Tokyo} in the European exchanges.
 Moreover, the share of the call auction in the European exchanges is increasing~\cite{derksen2022heavy,derksen2020effects}.

In this study, we  investigate the shape of the limit order books on the call auction, and see the differences among the stocks  with different attributes.
Moreover, we clarify the relation between the shape and the trading volume.

\section*{Results}
\subsection*{Fitting function of limit order book}

We use the data of the limit order books at the moment of the execution (9:00 A.M.) for all the stocks listed on the Tokyo Stock Exchange in 2022.
In order to analyze the shape of the limit order book for each stock, we aggregate the limit order books of all the days in the data for each stock.
Because the price levels and the order volumes vary day to day, we standardize them for aggregation.
The details of the data preparation are described in {\it Method}.

For the aggregated order distribution, we denote the number density of sell (Ask) limit orders at price $x$ as $n^A(x)$, and the number density of buy (Bid) limit orders at price $x$ as $n^B(x)$.
We denote  the cumulative number of sell  and buy limit orders as $\displaystyle N^A(x)=\int_0^{x}n^A(y)dy$ and  $\displaystyle N^B(x)=\int_x^{\infty}n^B(y)dy$, respectively.
Also,  the number of sell market orders and the buy market orders are denoted by $M^A$  and $M^B$, respectively.

The execution price $X$ and the trading volume $V$ are determined by the relation
\begin{equation}  
V=N^A(X)+M^A=N^B(X)+M^B.
\label{eq:cross_N}
\end{equation}
\begin{figure}[tb]
    \begin{minipage}[b]{0.45\linewidth}
        \centering
    \includegraphics[keepaspectratio, scale=0.5, clip, bb=22 23 410 340]
    {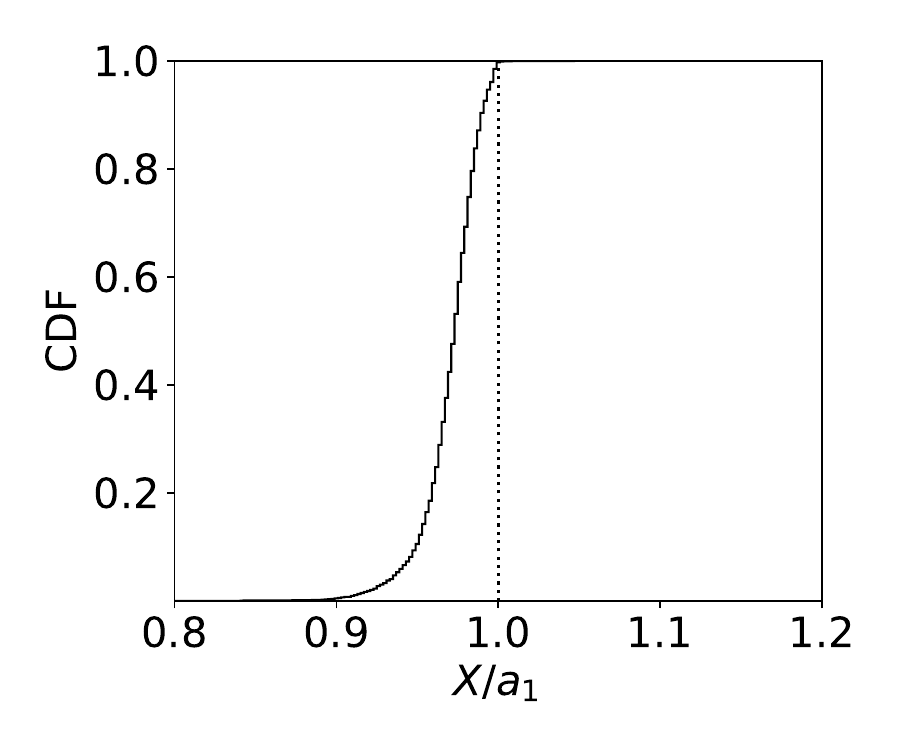}
    \end{minipage}
    \begin{minipage}[b]{0.45\linewidth}
        \centering
        \includegraphics[keepaspectratio, scale=0.5, clip, bb=22 23 410 340]
        {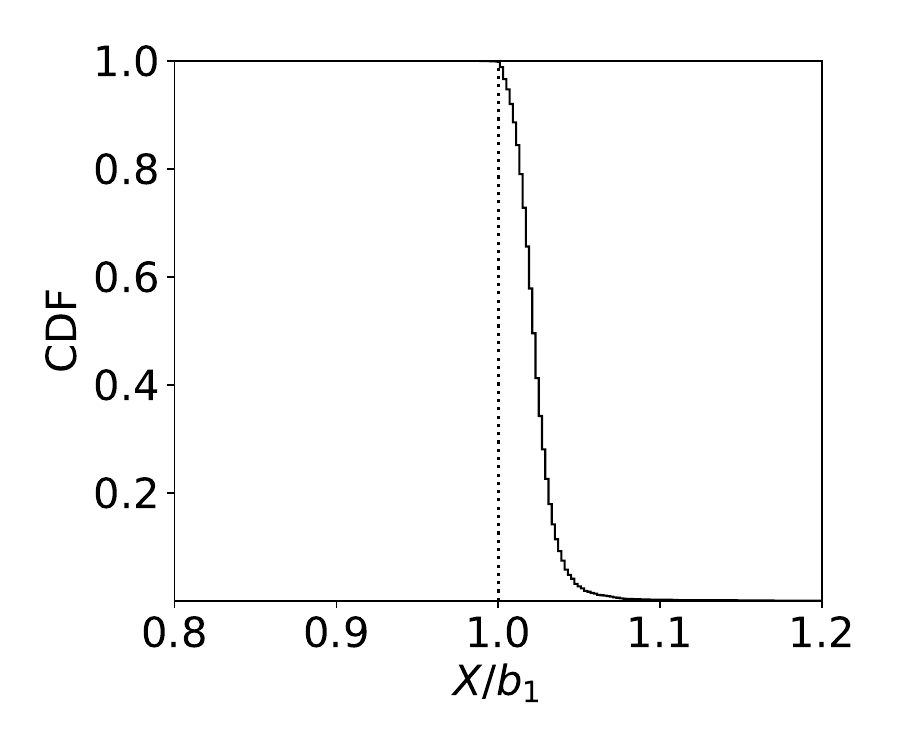}
    \end{minipage}
    \caption{
    The cumulative distribution of the relative position of the execution price $X$ to the first quartiles of sell and buy orders $a_1$ and $b_1$.
    Almost all execution prices are within the first quartiles.}
    \label{fig:hist_a025_b025_noFix}

    \centering
    \includegraphics[keepaspectratio, scale=0.5]{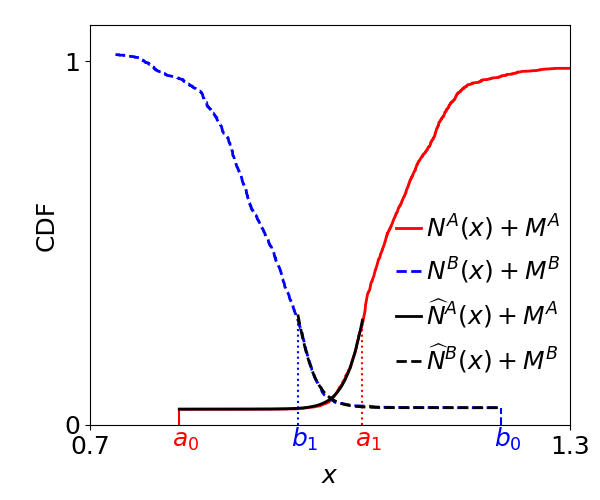}
    \caption{A typical shape of a limit order book with the fitting functions in Eq. (\ref{eq:fit}). 
    We treat sell market orders as the sell limit orders at the best ask $a_0$, so the number of sell orders  placed at $a_0$ includes sell market orders $M^A$, which appears
    Likewise, the cumulative total number of buy limit orders and buy market orders $N^B(x)+M^B$ starts from the best bid $b_0$, and the number of buy orders placed at $b_0$ includes buy market orders $M^B$.
    $N^{A}(x)$ is fitted by $\widehat{N}^{A}(x)$ in the range $[a_0,\:a_1]$, and $N^{B}(x)$ is fitted by $\widehat{N}^{B}(x)$ in the range $[b_1,\:b_0]$,
    where $\widehat{N}^{A}(x)$ and $\widehat{N}^{B}(x)$ are the fitting functions in Eq. (\ref{eq:fit}).
    }
    \label{fig:9780}
\end{figure}
As shown in Fig. \ref{fig:hist_a025_b025_noFix}, the execution price $X$ is in the range between the first quartile of sell limit orders, $a_1$ and the first quartile of buy limit orders, $b_1$.
Therefore, in order to keep the information to characterize the price and the volume, we fit $N^{A}(x)$ in the range $[a_0,\:a_1]$, and fit $N^{B}(x)$ in the range $[b_1,\:b_0]$ by functions introduced below, where $a_0$ is the price of the best ask and $b_0$ is the price of the best bid.
Fig. \ref{fig:9780} shows a typical shape of a limit order book.
We treat the sell market orders as the sell limit orders placed at $a_0$, and treat the buy market orders as the buy limit orders placed at $b_0$.
The red solid line represents the sell limit orders and market orders $N^A(x)+M^A$, and the blue dashed line represents the buy limit orders and market orders $N^B(x)+M^B$.
It is found that the cumulative orders $N^{A}(x)$ and $N^{B}(x)$ typically start from a slow rise and then steepens at a certain point like the sigmoid function, so we adopt the hyperbolic tangent function as the fitting function for its mathematical convenience.
Namely, we fit $N^{A}(x)$ and $N^{B}(x)$ as
\begin{equation}
    \widehat{N}^A(x)=\frac{N^A}{2}\left[\tanh\left(\frac{x-\alpha}{\omega^A}\right)+1\right],
    \:\:\:\:\:\:\:
    \widehat{N}^B(x)=\frac{N^B}{2}\left[\tanh\left(\frac{\beta-x}{\omega^B}\right)+1\right],
    \label{eq:fit}
\end{equation}
where $N^A=N^A(\infty)$ and  $N^B=N^B(0)$  are the total numbers of sell and buy limit orders.
$\omega^A$ and $\omega^B$ are the fitting parameters which correspond to the typical widths of the sell and buy orders.
$\alpha$ and $\beta$ represent the medians of the sell and buy fitting functions.
The example of the fit for the limit order book is shown in Fig. \ref{fig:9780} by the black lines.

\subsection*{Asymmetry in the width of the sell and buy limit order books}
\begin{figure}[h]
\vspace{1cm}
    \centering
 \includegraphics[keepaspectratio, scale=0.55, clip, bb=20 25 490 410]
    {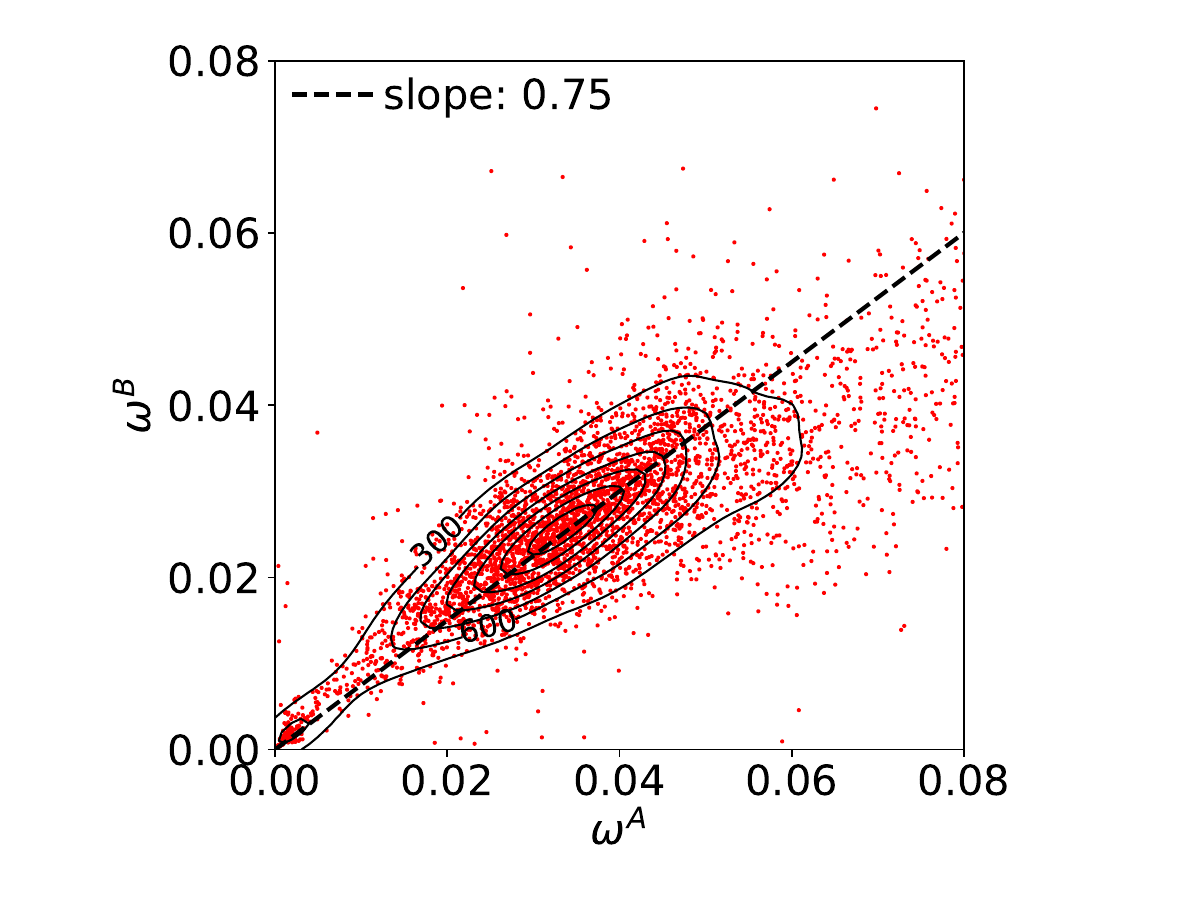}
  \caption{The asymmetry between the widths in sell and buy sides, $\omega^A$ and $\omega^B$.}
  \label{fig:contour_wA_wB_noFix}
\end{figure}

Using the fitting functions in Eq. (\ref{eq:fit}), we observe the relation between the typical widths of the sell and buy orders, $\omega^A$ and $\omega^B$.
Fig. \ref{fig:contour_wA_wB_noFix} shows the scatter plot of stocks in the $\omega^A$-$\omega^B$ plane with its density contours.
This indicates that $\omega^A$ and $\omega^B$ are correlated, with the clear asymmetry that $\omega^A$ tends to be larger than $\omega^B$.
In the continuous auction, the asymmetry in the sell-side and the buy-side has been reported for the number of orders~\cite{Noritake2022Impact}.
For the distributions of incoming orders, there seems to be no difference between the sell-side and the buy-side when the sell (buy) order prices are normalized by the best ask (bid)~\cite{mike2008empirical}.
The asymmetry in the shape of the limit order book we have found here is different from the above reported ones, which may be specific to the call auction.


\clearpage

\subsection*{Classification of stocks}
\begin{figure}[b]

  \begin{minipage}[b]{0.47\linewidth}
    \centering
    \includegraphics[keepaspectratio, scale=0.55, clip, bb=20 25 482 410]
    {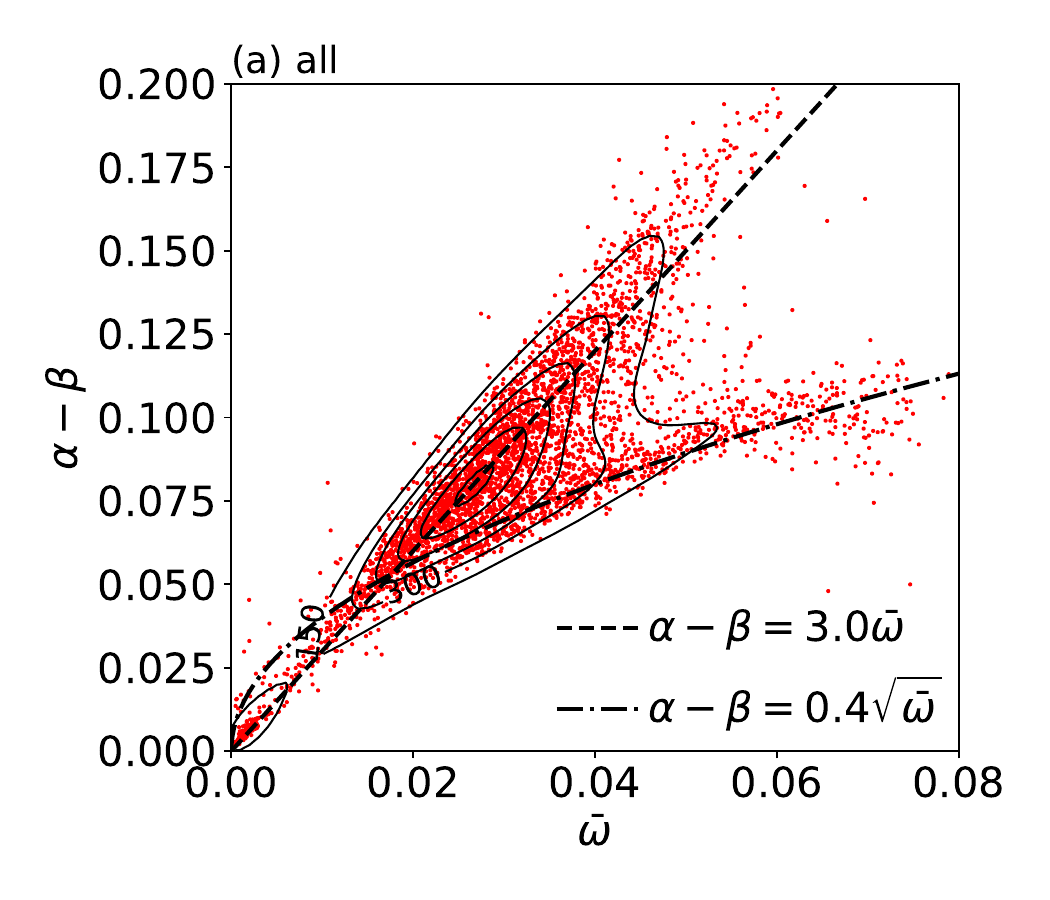}
  \end{minipage}
  \begin{minipage}[b]{0.47\linewidth}
    \centering
    \includegraphics[keepaspectratio, scale=0.47, clip, bb=20 25 482 410]
    {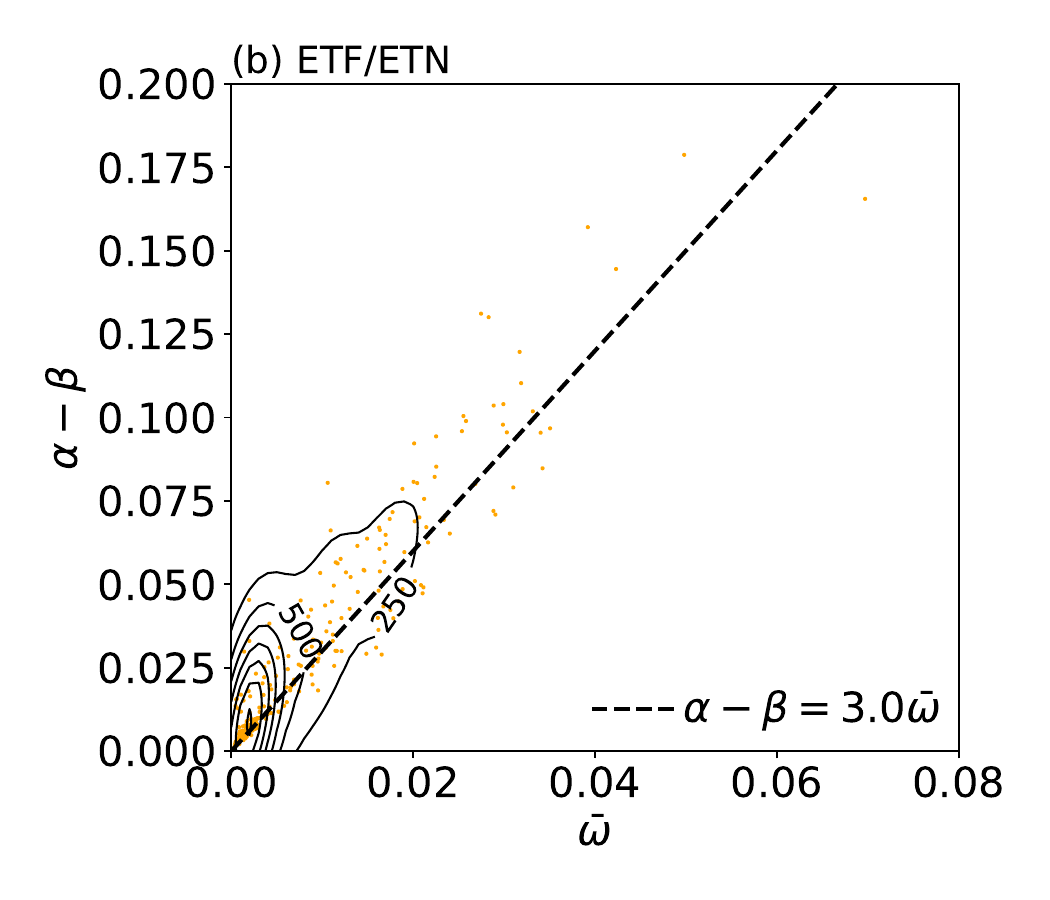}
  \end{minipage}
  \vspace{5mm}
    \begin{minipage}[b]{0.47\linewidth}
    \hspace{11mm}
    \includegraphics[keepaspectratio, scale=0.47, clip, bb=20 25 482 410]
    {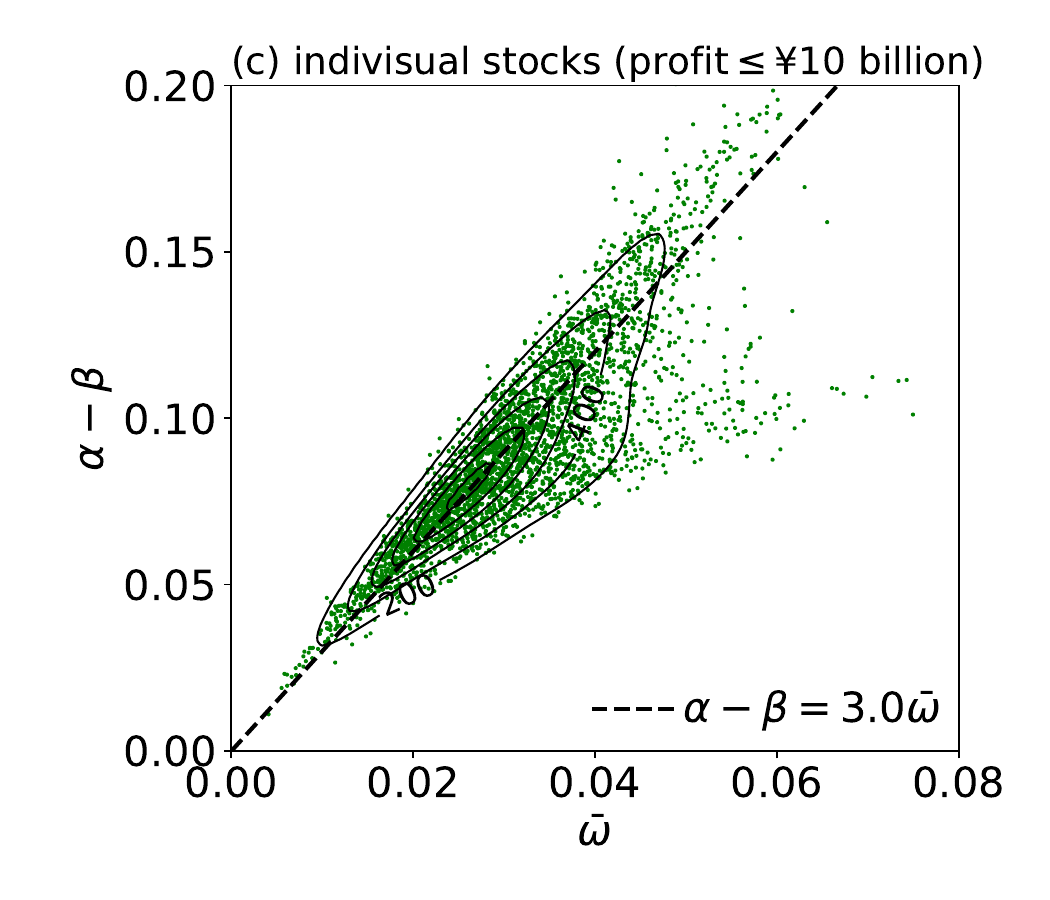}
  \end{minipage}
  \hspace{8.8mm}
   \begin{minipage}[b]{0.47\linewidth}
    \centering
    \includegraphics[keepaspectratio, scale=0.47, clip, bb=20 25 482 410]{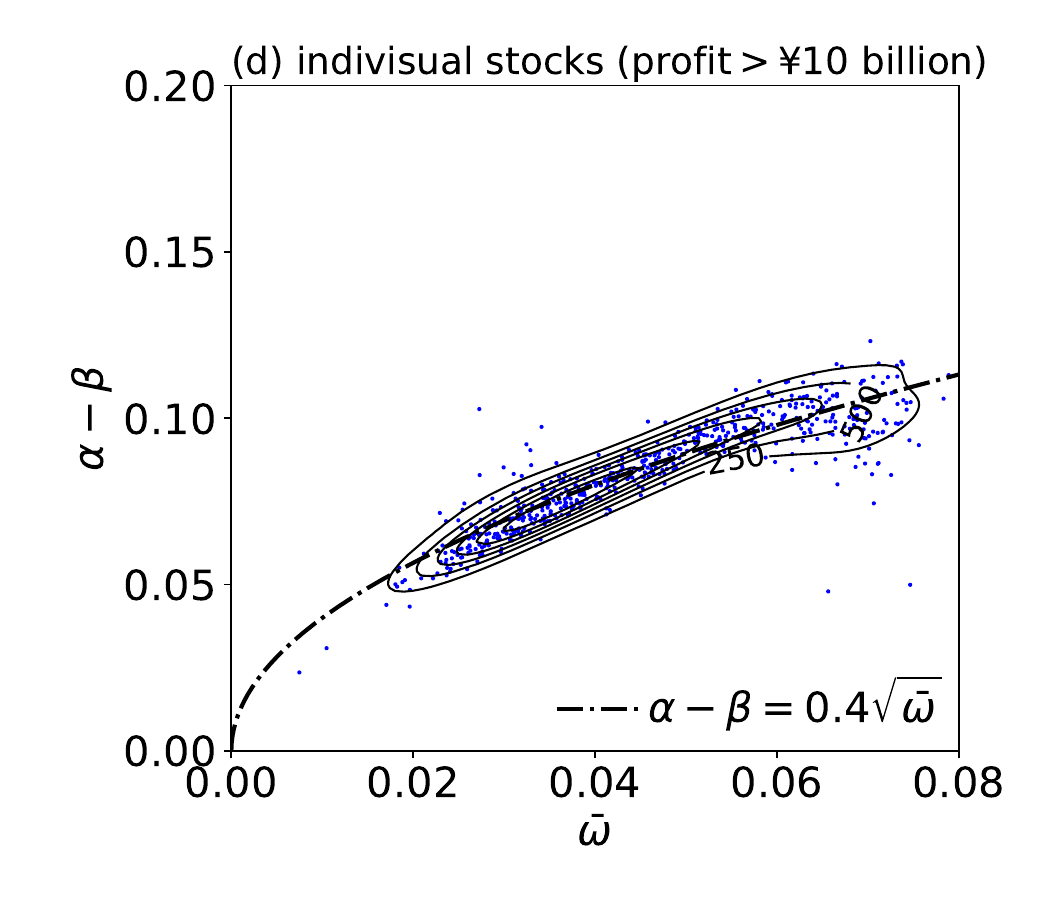}
    \end{minipage}
    \vspace{-7mm}
    \caption{
    Scatter plots of
    (a) All the stocks,
    (b) ETF/ETN, 
    (c) Individual stocks whose net profits are less than 10 billion yen, and 
    (d) Individual stocks whose net profits are more than 10 billion yen,
    on the plain of width ($\bar{\omega}$) and spread ($\alpha-\beta$).
    The density contours are plotted by the solid lines.
    The dashed line $(\alpha-\beta)/\bar{\omega}=3.0$ corresponds to the equal execution ratio ($\widehat{v}  = 4\%$) line, and the dash-dotted curve $\alpha-\beta=0.4\sqrt{\bar{\omega}}$ is for the eye guide. 
    }
    \label{fig:contour_w_a1-b1}

    \vspace{5mm}

    \begin{minipage}[b]{1\linewidth}
    \centering
    \includegraphics[keepaspectratio, scale=0.45]{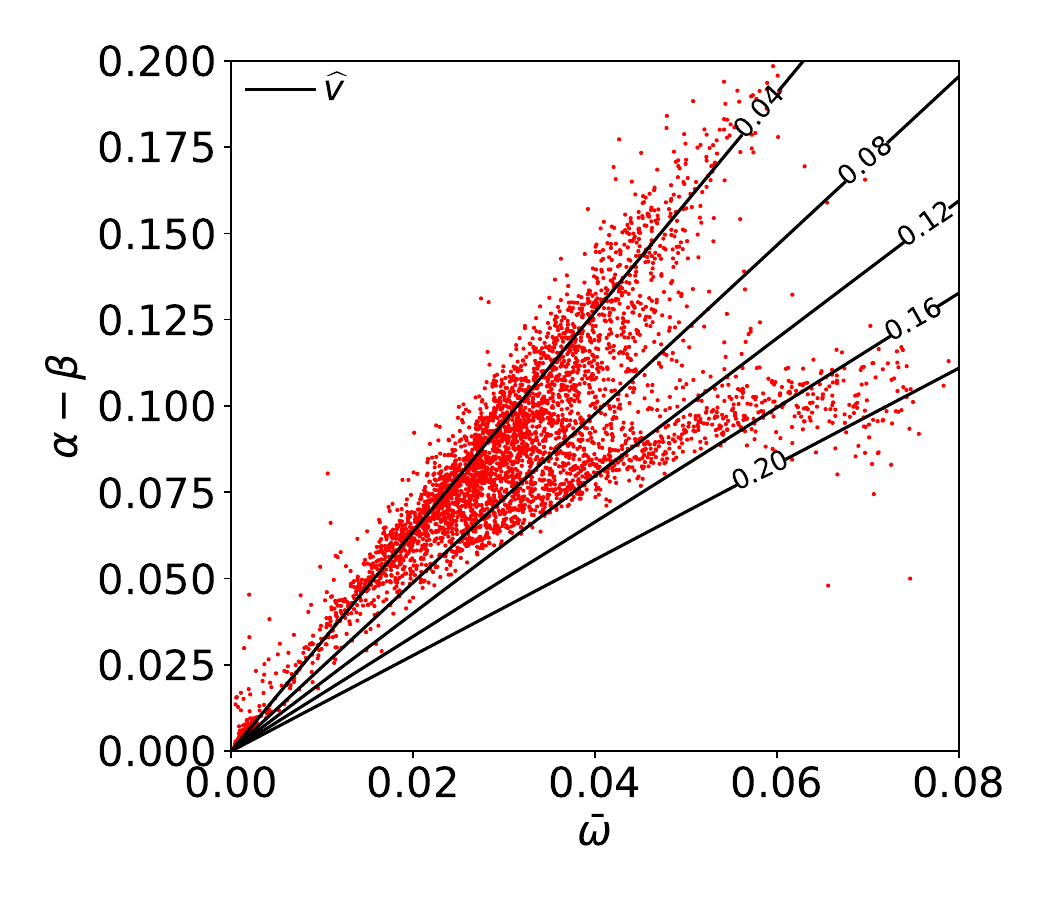}
    \end{minipage}
    \vspace{-10mm}
    \caption{ Equal execution ratio lines on the  $\bar{\omega}$ and $\alpha-\beta$ plain.
 The slope $3.0$ corresponds to the theoretical execution ratio  $\simeq 4\%$.
 Lines with gentler slopes correspond to higher execution ratios..
 }
    \label{fig:contour_w_a025-b025_theo_V_noFix}
        
\end{figure}

We now compare the distance between the sell orders and the buy orders to those widths, to capture the basic information of the limit order book of each stock.
We use $\bar{\omega}=(\omega^A + \omega^B)/2$ to characterize the width of the orders, and $\alpha-\beta$ to characterize the distance between the sell orders and the buy orders.
Fig.~\ref{fig:contour_w_a1-b1} (a) shows the distribution of stocks in the $\bar{\omega}$ and $\alpha-\beta$ plain.
This distribution can be classified into three clusters.
The first cluster (Cluster 1) is characterized by the small width and the small spread   ($\bar{\omega}< 0.01, \alpha-\beta <0.025$).
The second cluster (Cluster 2) is the group of stocks which are distributed around a  line: $\alpha-\beta=3\bar{\omega}$.
The other stocks are distributed below this line, forming the third cluster (Cluster 3).

We find that those clusters are strongly  correlated with  the types of stocks.
While most stocks in  Cluster 2 and Cluster 3 are individual stocks, 
most stocks in Cluster 1 are Exchange-Traded Funds (ETFs) and Exchange-Traded Notes (ETNs) (Fig.~\ref{fig:contour_w_a1-b1} (b)).
ETFs and ETNs are investment funds linked to an index composed of different stocks and other assets,
and therefore
the fluctuations of the prices tend to be small, merely because of the law of large numbers.
An interesting observation is that both Cluster 1 and Cluster 2 are
 distributed along the same line: $\alpha-\beta=3\bar{\omega}$.
This constant slope of the distribution corresponds to the constant execution ratio.
As is shown in the {\it Method}, the execution ratio $\widehat{v}$  for given fitting functions is determined only by the slope $ (\alpha-\beta)/\bar{\omega}$ as
\begin{equation}
    \widehat{v}=\frac{1}{2}\left[1-\tanh\left(\frac{\alpha-\beta}{2\bar{\omega}}\right)\right].
    \label{v}
\end{equation}
The lines for the constant execution ratio are shown in 
Fig.~\ref{fig:contour_w_a025-b025_theo_V_noFix}. 
The slope $ (\alpha-\beta)/\bar{\omega}=3.0$ corresponds to the execution ratio $\widehat{v}=4\%$.
The fact that Cluster 1 and Cluster 2 are distributed along the single line
suggests that self-organizing mechanism may work among traders. 
That is, traders have an empirical knowledge of the normal shape of the limit order book, i.e. the ratio of the spread to the width.
Based on this knowledge, they place orders, which result in the normal shape of the limit order book. 
As for Cluster 3,  they are distributed along a curve: $\alpha-\beta=0.4\sqrt{\bar{\omega}}$, which is below the line with slope $ (\alpha-\beta)/\bar{\omega}=3.0$.
For these stocks, the theoretical execution ratios $\widehat{v}$  are higher than $4\%$.
The origin of this non-linearity of the curve is not clear yet.
It turns out that
most of the individual stocks in Cluster 3 are separated from  Cluster 2 by whether the company's net profit in that year was more than 10 billion yen or not (Fig.~\ref{fig:contour_w_a1-b1} (c), (d)).

\clearpage

\section*{Summary and discussion}
Intensive studies have been made on the limit order books under the continuous auction.
On the other hand, there has been little research on the limit order books in the call auction.
In this study, we focused on the shape of the limit order books in the call auction at the moment of its execution.
The empirical findings we have discovered are following.

First, we fitted the sell (A) and buy (B) limit order books for each stock of call auction by the hyperbolic tangent function to obtain the typical widths $\omega^A, \omega^B$ and the medians $\alpha, \beta$ of them.
We find that the widths in the sell-side and the buy-side are correlated and show clear asymmetry that $\omega^A$ is larger than $\omega^B$.
This suggests that the valuations of sell traders are more diverse than buy traders.
Such asymmetry between the sell and buy limit order books has not been reported for the continuous auction.

Next, we examined the relation between the average width $\bar{\omega}=(\omega^A + \omega^B)/2$ and the median spread $\alpha-\beta$.
From that, we found that stocks are roughly classified into three clusters.
Cluster 1 is the group of stocks whose width and spread are small ($\bar{\omega}< 0.01, \alpha-\beta <0.025$).
Cluster 2 is the group of stocks which are distributed around the line: $\alpha-\beta=3\bar{\omega}$.
Cluster 3 is the group of stocks which are distributed below Cluster 2, forming a branch around a sub-linear curve: $\alpha-\beta=0.4\sqrt{\bar{\omega}}$.
Most stocks in Cluster 1 are ETFs and ETNs.
Most stocks in Cluster 2 and Cluster 3 are individual stocks. Cluster 2 is formed mainly by the stocks of the companies whose net profits are less than 10 billion yen, and the stocks in Cluster 3 tends to be the ones with net profits larger than 10 billion yen.
Moreover, we showed that the theoretical execution ratio $\widehat{v}$  is determined solely by the slope $(\alpha-\beta)/\bar{\omega}$.
The slope for Cluster 1 and Cluster 2, $(\alpha-\beta)/\bar{\omega}=3.0$, corresponds to the execution ratio $\widehat{v} = 4\%$.
The smaller slope for Cluster 3 corresponds to higher execution ratio: $\widehat{v}>4\%$.

The similar execution ratio of the majority of the stocks (Cluster 2) suggests the existence of an auto-regulation mechanism.
A possible scenario is described as the following.
When the spread is large relative to the width,
traders will place lower sell orders or higher buy orders because this may allow their orders to be executed prior to others, with keeping enough gain, i.e. the difference between the actual execution price and the trader's valuation price to that stock.
On the other hand, when the spread is small relative to the width,
traders will place higher sell orders or lower buy orders
to increase their gain with expecting to keep considerable
possibility of the execution.
In the former case, the execution ratio increases, and in the latter case, the execution ratio decreases.
Therefore, as a result of this 
mechanism, the execution ratio tends to be auto-regulated around a constant value.
Such mechanism is absent in the conventional models of continuous auction with limit order book structure, such as the Maslov model, which have successfully explained essential characteristics of the real market dynamics such as the price fluctuations~\cite{maslov2000simple,slanina2001mean}.
Since the trader's character to seek higher execution ratio and gain should be also relevant in the continuous auction, introducing such mechanism into the continuous auction models is an important direction for future study.
The relation between trading volume and gain has been also studied by a simpler mathematical model~\cite{nagumo2018Impact,nagumo2022analytical}, which is again without any auto-regulatory mechanism.
Applying such mechanism into this framework will improve the understanding of market, such as the merit of the increase of the number of participants, clarifying the quantity which is subject to the maximization in the trader's strategy, and so on.

The finding that Cluster 3 which mainly consists of stocks with profits more than 10 billion yen is distributed below the majority (Cluster 2) could be interpreted in terms of traders' risk appetite, as explained in the previous study~\cite{agrawal2004bid}.
That is, for such 
favored stocks, traders' appetite is mostly on trading that stocks (i.e. the higher execution probability) and not so much on the gain.
However, we do not have good understanding for why Cluster 3 appears as a distinct branch apart from Cluster 2, rather than as a continuously distributed tail.
Also, the origin of the sub-linear shape ($\alpha-\beta=0.4\sqrt{\bar{\omega}}$) of Cluster 3 remains unknown.

Clusters 1, 2, and 3 have been discovered on the plain of width ($\bar{\omega}$) and median spread ($\alpha-\beta$).
However,
if we substitute the median spread by conventional BBO spread ($a_0-b_0$), such clustering structure disappears (see {\it Supplementary Information}).
This illustrates the importance of taking the shape of the limit order book apart from best ask (bid) to better capture the state of the stock market.

Our finding on the shape of limit order books could be applied
to the study on market impact.
It is well known that the market impact in the continuous auction is proportional to the square root of the order volume.
This fact has been explained by the latent order book model, which is based on the linearity of the limit order book near the execution price~\cite{toth2011anomalous}.
From the curvature of the shape of limit order books which we have found in this study, the market impact in the call auction can be evaluated more precisely.
As another practical application, we can consider the daily monitoring of $\bar{\omega}$ and $\alpha-\beta$ of limit order books.
Suppose that $\bar{\omega}$ and $\alpha-\beta$ for a stock is found in a different position than usual.
That helps stock exchanges to detect erroneous orders or illegal activities such as insider trading.
For traders, unusual position of $\bar{\omega}$ and $\alpha-\beta$ may indicate unexpected changes in the company performance such as the profit, and contribute to their investment decisions. 

\section*{Method}
\subsection*{Data set}

We use the data of limit order books at the time of  the call auction at 9:00 A.M. for 244 business days from January 4 to December 30, 2022 for all the listed stocks on the Tokyo Stock Exchange (4,024 stocks per day on average).
Stocks for which the number of orders is so small that the daily best ask, best bid and the first quartiles cannot be defined are to be excluded.
In order to examine the shape of the limit order book for each stock, we aggregate the limit orders of all the days in the data for each stock.

\subsection*{Data preparation}
The order volumes vary day to day.
Sometimes artificial events such as stock split are conducted by the Tokyo Stock Exchange.
Therefore, when we aggregate the limit order books, the weight of the order volume for each day is set to be constant.


Also, the stock price levels varies from day to day, and  according to these differences, the position of the limit order book varies from day to day.
Therefore, we define a reference price for each day for each stock, and calculate order prices divided by the reference price.
We aggregate the limit order books based on the relative price.
This concept of the relative price is used in a previous study~\cite{mike2008empirical}, for instance.
As for the definition of the reference price, the execution price of the call auction is one candidate.
However, the execution price cannot be defined when the sell and buy limit order books are not overlapped.
Next idea of the definition of the reference price shall be the middle price of the best ask and the best bid.
However, for many cases of call auctions, the best ask and the best bid become the lower and the upper limit price, which are artificially set by the Tokyo Stock Exchange.
Therefore, the middle price of the best ask and the best bid does not reflect the stock price level of the day.
Thus, we define the reference price by the average of the first quartile of the sell limit order book and the first quartile of the buy limit order book.

\subsection*{Theoretical execution price and execution ratio}

As Eq. (\ref{eq:cross_N}), the execution is conducted at the point where the cumulative numbers of sell and buy orders cross.
By using the fitting functions $\widehat{N}^A(x)$ and $\widehat{N}^B(x)$ obtained in Eq. (\ref{eq:fit}), the execution price $\widehat{X}$ and the trading volume $\widehat{V}$ shall satisfy the following relation
\begin{equation}
\widehat{V}=\widehat{N}^A(\widehat{X})+M^A=\widehat{N}^B(\widehat{X})+M^B.
    \label{eq:cross}
\end{equation}
Under the approximations that $N^A=N^B $, $M^A=M^B $, and $\omega^A=\omega^B$,
we  obtain the execution price $\widehat{X}$ and the execution ratio $\widehat{v}=(\widehat{V}-M^A)/N^A$ as follows.
\begin{equation}
    \widehat{X}=\frac{\alpha+\beta}{2}, \:\:\:\:\:\:
    \widehat{v}=\frac{1}{2}\left[1-\tanh\left(\frac{\alpha-\beta}{2\omega}\right)\right].
    \label{eq:v}
\end{equation}

\section*{Data availability statement}
The data that support the findings of this study are available from Tokyo Stock Exchange, Inc. but restrictions apply to the availability of these data, which were used under the non-disclosure agreement for the current study, and so are not publicly available. However, in some cases, the processed data, not the raw data may be shared by the authors upon reasonable request and with the permission of Japan Exchange Group, Inc. (the parent company of Tokyo Stock Exchange, Inc.).

\bibliographystyle{splncs04}


\begin{thebibliography}{10}
\urlstyle{rm}
\expandafter\ifx\csname url\endcsname\relax
  \def\url#1{\texttt{#1}}\fi
\expandafter\ifx\csname urlprefix\endcsname\relax\def\urlprefix{URL }\fi
\expandafter\ifx\csname doiprefix\endcsname\relax\def\doiprefix{DOI: }\fi
\providecommand{\bibinfo}[2]{#2}
\providecommand{\eprint}[2][]{\url{#2}}

\bibitem{chakraborti2011econophysics1}
\bibinfo{author}{Chakraborti, A.}, \bibinfo{author}{Toke, I.~M.}, \bibinfo{author}{Patriarca, M.} \& \bibinfo{author}{Abergel, F.}
\newblock \bibinfo{journal}{\bibinfo{title}{Econophysics review: I. empirical facts}}.
\newblock {\emph{\JournalTitle{Quantitative Finance}}} \textbf{\bibinfo{volume}{11}}, \bibinfo{pages}{991--1012} (\bibinfo{year}{2011}).

\bibitem{chakraborti2011econophysics2}
\bibinfo{author}{Chakraborti, A.}, \bibinfo{author}{Toke, I.~M.}, \bibinfo{author}{Patriarca, M.} \& \bibinfo{author}{Abergel, F.}
\newblock \bibinfo{journal}{\bibinfo{title}{Econophysics review: {II}. agent-based models}}.
\newblock {\emph{\JournalTitle{Quantitative Finance}}} \textbf{\bibinfo{volume}{11}}, \bibinfo{pages}{1013--1041} (\bibinfo{year}{2011}).

\bibitem{bouchaud2009markets}
\bibinfo{author}{Bouchaud, J.-P.}, \bibinfo{author}{Farmer, J.~D.} \& \bibinfo{author}{Lillo, F.}
\newblock \bibinfo{title}{How markets slowly digest changes in supply and demand}.
\newblock In \emph{\bibinfo{booktitle}{Handbook of financial markets: dynamics and evolution}}, \bibinfo{pages}{57--160} (\bibinfo{publisher}{Elsevier}, \bibinfo{year}{2009}).

\bibitem{bouchaud2018trades}
\bibinfo{author}{Bouchaud, J.-P.}, \bibinfo{author}{Bonart, J.}, \bibinfo{author}{Donier, J.} \& \bibinfo{author}{Gould, M.}
\newblock \emph{\bibinfo{title}{Trades, quotes and prices: financial markets under the microscope}} (\bibinfo{publisher}{Cambridge University Press}, \bibinfo{year}{2018}).

\bibitem{Abergel2016Limit}
\bibinfo{author}{Abergel, F.}, \bibinfo{author}{Anane, M.}, \bibinfo{author}{Chakraborti, A.}, \bibinfo{author}{Jedidi, A.} \& \bibinfo{author}{Muni~Toke, I.}
\newblock \emph{\bibinfo{title}{Limit Order Books}}.
\newblock Physics of Society: Econophysics and Sociophysics (\bibinfo{publisher}{Cambridge University Press}, \bibinfo{year}{2016}).

\bibitem{gould2013limit}
\bibinfo{author}{Gould, M.~D.} \emph{et~al.}
\newblock \bibinfo{journal}{\bibinfo{title}{Limit order books}}.
\newblock {\emph{\JournalTitle{Quantitative Finance}}} \textbf{\bibinfo{volume}{13}}, \bibinfo{pages}{1709--1742} (\bibinfo{year}{2013}).

\bibitem{kyle1985continuous}
\bibinfo{author}{Kyle, A.~S.}
\newblock \bibinfo{journal}{\bibinfo{title}{Continuous auctions and insider trading}}.
\newblock {\emph{\JournalTitle{Econometrica: Journal of the Econometric Society}}} \bibinfo{pages}{1315--1335} (\bibinfo{year}{1985}).

\bibitem{agrawal2004bid}
\bibinfo{author}{Agrawal, V.}, \bibinfo{author}{Kothare, M.}, \bibinfo{author}{Rao, R.~K.} \& \bibinfo{author}{Wadhwa, P.}
\newblock \bibinfo{journal}{\bibinfo{title}{Bid-ask spreads, informed investors, and the firm’s financial condition}}.
\newblock {\emph{\JournalTitle{The Quarterly Review of Economics and Finance}}} \textbf{\bibinfo{volume}{44}}, \bibinfo{pages}{58--76} (\bibinfo{year}{2004}).

\bibitem{naes2006order}
\bibinfo{author}{N{\ae}s, R.} \& \bibinfo{author}{Skjeltorp, J.~A.}
\newblock \bibinfo{journal}{\bibinfo{title}{Order book characteristics and the volume--volatility relation: Empirical evidence from a limit order market}}.
\newblock {\emph{\JournalTitle{Journal of Financial Markets}}} \textbf{\bibinfo{volume}{9}}, \bibinfo{pages}{408--432} (\bibinfo{year}{2006}).

\bibitem{lehmann1994trading}
\bibinfo{author}{Lehmann, B.~N.} \& \bibinfo{author}{Modest, D.~M.}
\newblock \bibinfo{journal}{\bibinfo{title}{Trading and liquidity on the tokyo stock exchange: A bird's eye view}}.
\newblock {\emph{\JournalTitle{The Journal of Finance}}} \textbf{\bibinfo{volume}{49}}, \bibinfo{pages}{951--984} (\bibinfo{year}{1994}).

\bibitem{Ohta2008Tokyo}
\bibinfo{author}{Ohta, W.}
\newblock \bibinfo{journal}{\bibinfo{title}{Stock trading on the tokyo stock exchange from 2001 to 2003 (in {Japanese})}}.
\newblock {\emph{\JournalTitle{Osaka University Knowledge Archive}}} \textbf{\bibinfo{volume}{57}}, \bibinfo{pages}{242--262} (\bibinfo{year}{2008}).

\bibitem{derksen2022heavy}
\bibinfo{author}{Derksen, M.}, \bibinfo{author}{Kleijn, B.} \& \bibinfo{author}{De~Vilder, R.}
\newblock \bibinfo{journal}{\bibinfo{title}{Heavy tailed distributions in closing auctions}}.
\newblock {\emph{\JournalTitle{Physica A: Statistical Mechanics and its Applications}}} \textbf{\bibinfo{volume}{593}}, \bibinfo{pages}{126959} (\bibinfo{year}{2022}).

\bibitem{derksen2020effects}
\bibinfo{author}{Derksen, M.}, \bibinfo{author}{Kleijn, B.} \& \bibinfo{author}{De~Vilder, R.}
\newblock \bibinfo{journal}{\bibinfo{title}{Effects of mifid ii on stock price formation}}.
\newblock {\emph{\JournalTitle{arXiv preprint arXiv:2003.10353}}}  (\bibinfo{year}{2020}).

\bibitem{Noritake2022Impact}
\bibinfo{author}{Noritake, Y.}, \bibinfo{author}{Hemmi, R.}, \bibinfo{author}{Nagumo, S.}, \bibinfo{author}{Mizuta, T.} \& \bibinfo{author}{Izumi, K.}
\newblock \bibinfo{journal}{\bibinfo{title}{Analysis of short side market inefficiencies using artificial market model}}.
\newblock {\emph{\JournalTitle{JPX working paper}}} \textbf{\bibinfo{volume}{38}} (\bibinfo{year}{2022}).

\bibitem{mike2008empirical}
\bibinfo{author}{Mike, S.} \& \bibinfo{author}{Farmer, J.~D.}
\newblock \bibinfo{journal}{\bibinfo{title}{An empirical behavioral model of liquidity and volatility}}.
\newblock {\emph{\JournalTitle{Journal of Economic Dynamics and Control}}} \textbf{\bibinfo{volume}{32}}, \bibinfo{pages}{200--234} (\bibinfo{year}{2008}).

\bibitem{maslov2000simple}
\bibinfo{author}{Maslov, S.}
\newblock \bibinfo{journal}{\bibinfo{title}{Simple model of a limit order-driven market}}.
\newblock {\emph{\JournalTitle{Physica A: Statistical Mechanics and its Applications}}} \textbf{\bibinfo{volume}{278}}, \bibinfo{pages}{571--578} (\bibinfo{year}{2000}).

\bibitem{slanina2001mean}
\bibinfo{author}{Slanina, F.}
\newblock \bibinfo{journal}{\bibinfo{title}{Mean-field approximation for a limit order driven market model}}.
\newblock {\emph{\JournalTitle{Physical Review E}}} \textbf{\bibinfo{volume}{64}}, \bibinfo{pages}{056136} (\bibinfo{year}{2001}).

\bibitem{nagumo2018Impact}
\bibinfo{author}{Nagumo, S.} \& \bibinfo{author}{Ichiki, S.}
\newblock \bibinfo{journal}{\bibinfo{title}{Impact on utility of traders by improvement of liquidity in stock secondary markets (in {Japanese})}}.
\newblock {\emph{\JournalTitle{JPX working paper}}} \textbf{\bibinfo{volume}{24}} (\bibinfo{year}{2018}).

\bibitem{nagumo2022analytical}
\bibinfo{author}{Nagumo, S.}, \bibinfo{author}{Ichiki, S.} \& \bibinfo{author}{Shimada, T.}
\newblock \bibinfo{journal}{\bibinfo{title}{{Analytical and simulational approaches to the relation between the stock market liquidity and the traders' utility}}}.
\newblock {\emph{\JournalTitle{Artificial Life and Robotics}}} \textbf{\bibinfo{volume}{27}}, \bibinfo{pages}{691--697} (\bibinfo{year}{2022}).

\bibitem{toth2011anomalous}
\bibinfo{author}{T{\'o}th, B.} \emph{et~al.}
\newblock \bibinfo{journal}{\bibinfo{title}{Anomalous price impact and the critical nature of liquidity in financial markets}}.
\newblock {\emph{\JournalTitle{Physical Review X}}} \textbf{\bibinfo{volume}{1}}, \bibinfo{pages}{021006} (\bibinfo{year}{2011}).

\end{thebibliography}

\section*{Acknowledgements}
We greatly appreciate the data provision by Tokyo Stock Exchange, Inc.

\section*{Author contributions}
S.N. and T.S. conceived the study. S.N. analyzed the data. S.N. and T.S. discussed the results and wrote the manuscript.

\section*{Competing interests}
The authors declare no competing interests.

\section*{Disclaimer}
The views expressed in this paper are those of the authors and do not represent the views of the organizations to which they belong.


\end{document}